# Scanning Tunneling Microscopy of Gate Tunable Topological Insulator Bi$_2$Se$_3$ Thin Films

Tong Zhang[1,2], Niv Levy[1], Jeonghoon Ha[1,2,3], Young Kuk[3], and Joseph A. Stroscio[1]*

[1]Center for Nanoscale Science and Technology, NIST, Gaithersburg, MD 20899, USA
[2]Maryland NanoCenter, University of Maryland, College Park, MD 20742, USA
[3] Department of Physics and Astronomy, Seoul National University, Seoul, 151-747, Korea

Electrical field control of the carrier density of topological insulators (TI) has greatly expanded the possible practical use of these materials. However, the combination of low temperature local probe studies and a gate tunable TI device remains challenging. We have overcome this limitation by scanning tunneling microscopy and spectroscopy measurements on *in-situ* molecular beam epitaxy growth of Bi$_2$Se$_3$ films on SrTiO$_3$ substrates with pre-patterned electrodes. Using this gating method, we are able to shift the Fermi level of the top surface states by ≈250 meV on a 3 nm thick Bi$_2$Se$_3$ device. We report field effect studies of the surface state dispersion, band gap, and electronic structure at the Fermi level.

Three-dimensional (3D) topological insulators (TI) are a novel state of matter which has a bulk band gap but topologically protected metallic surface states [1,2]. Angle resolved photoemission (ARPES) [3–6] and scanning tunneling microscopy (STM) [7–10] have confirmed the existence of these surface states in various compounds. These protected surface states are helical Dirac fermions which are predicted to host many striking quantum phenomena [1,2]. However, in order to fully utilize the unique properties of these surface states, the Fermi level ($E_F$) needs to be close to the Dirac Point ($E_D$) and tunable across it. Unfortunately, three-dimensional TI, like $Bi_2Se_3$ and related materials, are usually heavily doped narrow gap semiconductors, with the Fermi level away from Dirac point. While bulk and surface chemical doping have been used to tune $E_F$, it is preferable to tune the carrier density using a gate induced electric field, as recently demonstrated in transport experiments [11–16]. However, the combination of local probe studies of the density of states using STM and a gate tunable TI device remains challenging, mostly due to the environmental sensitivity of TI materials [17]. Unlike graphene, *ex-situ* fabrication and processing will significantly degrade the surfaces of TI materials, making them inaccessible to STM. In addition to simply shifting $E_F$, the electric field is capable of altering the surface band structure of ultrathin TI films, where the top and bottom surface states are coupled, and a hybridization gap opens at the Dirac point [18–21]. As a consequence, it might be possible to observe a topological phase transition in these systems by applying an electric field perpendicular to the plane of the film, as suggested by recent calculations [21].

In this work, we demonstrate the fabrication of gate-tunable 3D TI devices that are suitable for STM studies. Thin $Bi_2Se_3$ films are epitaxially grown on $SrTiO_3$ (STO) (111) substrates pre-patterned with Pt electrodes, which are mounted on specially designed sample holders (Fig. 1). The preparation of the STO substrates is crucial for good quality film growth and a gate insulator that maintains high resistance. The STO (111) substrates were cleaned by the method described in Ref. [22]. STO pieces, $(3\times4\times0.1)$ mm$^3$, were immersed in hot deionized water (70 °C) with ultrasonic agitation for 30 min. The pieces were then annealed in a tube furnace at 1000 °C for 1 h under a pure $O_2$ atmosphere. After this two-step treatment, the STO surface is completely cleaned and ambient atomic force microscopy (AFM) images show flat and regular terraces (Fig. 2(a)). Additionally this treatment results in highly insulating STO, which is crucial for using it as a gate dielectric. After cleaning, two rectangular Pt electrodes (50 nm thick) were deposited on the top surface of the STO substrate to ensure good electrical contacts to the TI film. Another Pt electrode, serving as a back gate, was deposited on the bottom surface of STO. The pre-patterned STO piece was mounted into an $Al_2O_3$ based sample holder which has multiple tungsten clips for sample wiring and a tungsten spring clip to hold the sample, which also doubles as a gate contact (Fig. 1). The whole device is fully compatible with ultra-high vacuum and can be degassed to 600 °C. Prior to film growth, the STO and sample holder were degassed for 30 min at 500 °C to remove adsorbed gases. The reflection high-energy electron diffraction (RHEED) pattern gave clean $3\times3$ surface reconstruction patterns of STO (111) (Fig. 2(b)). Growth of $Bi_2Se_3$ films was carried out by co-evaporating pure elemental (99.9999%) Bi and Se from Knudsen cells, using a 1:10 flux ratio (Bi:Se) to reduce Se vacancies. The STO substrates were kept at 250 °C during growth. RHEED patterns (Fig.2(c)) measured during growth show 1×1 streaks, an indication of good crystal quality. The samples were then transferred *in-situ* into a STM directly after growth, which avoids any *ex-situ* post processing. The experiments were performed in a custom designed STM operating at 5 K, which is connected to the MBE systems [23]. Due to the large dielectric constant of STO at low temperatures ($\approx 10^4$), we were able to change the carrier density on the order of $10^{13}$ cm$^{-2}$ at a gate voltage of 100 V. Note, this corresponds to a similar density change in graphene obtainable with a gate insulator of 300 nm thick $SiO_2$ at 100 V, but we are able to reach similar densities with 100 µm thick STO crystal due to the large dielectric constant.

$Bi_2Se_3$ has a layered structure consisting of Se-Bi-Se-Bi-Se quintuple layers (QL), where the bonding between adjacent quintuple layers is weak [24]. In this paper we focus on three QL thick $Bi_2Se_3$ samples, since gating is more effective on thinner films. Figure 2(d) shows the STM topography of a nominal 3QL thick film with flat terraces and 1 nm step heights, corresponding to one QL. Figure 3(a) shows the two-terminal resistance measured during sample cool-down, which displays an insulating behavior with resistance increasing about two orders of magnitude between room temperature and 5 K. Similar insulating behavior has always been seen in ultrathin $Bi_2Se_3$ films, which could be due to strong interactions [25], or Anderson localization [26,27]. To test the gating ability of the TI/STO device, we measured the 2-terminal resistance through the 3QL film vs. gate voltage (Fig. 3(b)). The resistance increases at negative gate voltage as expected for an *n*-type doped grown film. However, a maximum is not observed which indicates that we were unable to place the Dirac point at the Fermi level even with -300 V applied to the gate electrode (Fig. 3(b)).

The gating effect was then locally characterized by scanning tunneling spectroscopy. The tunneling conductance *dI/dV* is measured by standard lock-in techniques and is shown in Fig 4(a) for different gate voltages $V_G$. At $V_G$ = 0 V, there is a "V" shaped structure in the *dI/dV* spectra with a minimum at a sample bias $V_B \approx$ -0.4 V, which is typical for $Bi_2Se_3$ and was presumably attributed to the surface Dirac cone. [9,28]. An additional feature seen in the spectra is a peak at $V_B \approx$ -0.5 V at $V_G$ = 0 V, which most possibly comes from a quantum well (QW) state of the thin film, as seen in similar 3D TIs [29], followed by a strong upturn due to the onset of the valence band top. At positive bias a kink in the spectrum is observed at $E_F$ ($V_B$= 0V) followed by an upturn.

Application of a gate electric field causes a shift of the spectra as observed in Fig. 4(a). A downward sweep of gate voltage is accompanied by the Dirac point shifting to higher energies relative to the Fermi level at 0 V, which is a direct signature of tuning of the carrier density by the applied electric field. To quantify the gating efficiency we plot the peak position of the QW peak vs. gate voltage in Fig. 4(b)). A linear fit to the data in Fig. 4(b)) gives a gating efficiency of (1.2±0.1) mV/V [30]. At 100 V applied gate potential, the resulting Dirac point shift is 120 meV, which gives a density change of surface carriers of $\approx$ 0.4x$10^{13}$ cm$^{-2}$ (assuming a linear Dirac spectrum with velocity 5x$10^5$ m/s). Assuming a simple capacitor model we expect a gate capacitance of $C_G = \varepsilon_0 \varepsilon / t = 620$ μF/m$^2$, where $\varepsilon_0$ is the permittivity of free space, *t* is the thickness (0.1 mm) and $\varepsilon \approx 7000$ is the dielectric constant of STO [31]. This gate capacitance should induce a charge density of $\approx 4\times 10^{13}$ cm$^{-2}$ at $V_G$ = 100 V. However, we observe an induced density that is ten times less, and one that is not sufficient to overcome the initial film doping and place the Dirac point at $E_F$ at $V_G$ = -100 V. This accounts for the lack of a maximum in the resistance vs. gate voltage plot in Fig. 3(b). We attribute the reduced gating efficiency of the top surface states, which decreases with increasing thickness, as due to screening by the degenerately doped bulk Bi$_2$Se$_3$ film, which screens the electric field at the top surface. As STM is mostly sensitive to top surface, the shift of the bottom surface's Dirac point may be larger than that of the top surface. The high initial doping of the film is probably due to defects produced in growth, which is common in MBE growth of Bi$_2$Se$_3$ films as well as bulk synthesized crystals. The mixed compound Bi$_2$Te$_2$Se has been found to be a more insulating material [32], and hence may respond better to our gating scheme.

In a 3QL Bi$_2$Se$_3$ film, a hybridized gap is expected to open at the Dirac point due to the coupling of top and bottom surface states [18], which has been observed in ARPES measurements [6]. Evidence of a gap opening can be seen in the spectra in Fig. 4(a), where a gap is seen between the QW state peak and the broad hump to the right of the Dirac minimum, as outlined by the dashed lines. The gap size is about 160 meV, and roughly matches the ARPES data [6]. However, the gap is not well defined when the spectra shifts to larger negative energies and higher doping.

In addition to gate dependent features in the spectra, we observe a gate independent kink in the tunneling spectra at $E_F$ (Fig. 4(a)). Similar DOS kinks around $E_F$ have been widely observed in $Bi_2Se_3$, for both thin film and bulk samples [9,10], and were typically considered to be the onset of the conduction band. However, the gate independence of the feature shown here contradicts that explanation and strongly suggests it stems from a different mechanism, such as the many body effects in 2D Dirac systems [33,34]. One possible mechanism would involve the surface plasmon mode which has been predicted to exist in graphene and TI surfaces [33]. Below $E_F$, the electron-plasmon interaction can lead to reconstruction of the Dirac cone by forming a plasmaron band, which gives a sudden DOS change at Fermi level [33]. This mechanism which depends on interactions far from the Dirac point may remain intact despite the opening of a gap around it [19], and may even be enhanced in ultrathin film geometries [25].

To further explore the gating effects on the surface state band structure, we studied the quasi-particle interference (QPI) pattern at different gate voltages. Fig. 5, (a)-(e) and (f)-(j) show the *dI/dV* mapping of the energy resolved local density of states (LDOS) at a step edge in Fig. 2(d) at $V_G = 0$ and $V_G = -100$ V, respectively. Within the bias range of 0.2 V to 0.6 V, standing wave patterns at the step edges are clearly observed, while below 0.2 V we could not observe a clear interference pattern. At biases close to the Dirac point, scattering is weakened due to topological protection, despite the opening of a gap at the Dirac point [19]. The data in Fig. 5 covers a bias range that is far from the Dirac point ($E_D$ is at $V_B = -0.4$ V for $V_G = 0$ V), and the constant energy contours of the surface states are hexagonally warped at these energies [35,36], which enhances the scattering.

The QPI periods in Fig. 5 are seen to depend on gate voltage. We analyze these patterns by considering the scattering geometry and Fermi surface contours in Fig. 6. Step edges occur along the closed packed directions in the lattice (Fig. 6(a)), and therefore the scattering in Fig. 5 across the steps in Fig. 2(d) represents scattering in the $\Gamma-M$ direction (Fig. 6(b)). The energy contours for $Bi_2Se_3$ can be estimated from the energy dispersion which describes the warping given by [35],

$$E(k)_\pm = \pm\sqrt{(vk)^2 + (\lambda k^3 \cos 3\theta_k)^2} \qquad (1)$$

where $v$ is the Dirac velocity, $\lambda$ is the warping parameter, and $\theta_k = \tan^{-1}(k_y/k_x)$ is the azimuthal angle of the momentum with respect to x axis $(\Gamma-K)$. The energy contours using Eq. 1 for $Bi_2Se_3$ with $\lambda$=128 eV Å$^3$ and v=3.55 eV Å [36] are shown in Fig. 6(c). Above 0.3 eV relative to the Dirac point the contours deviate from circular symmetry. The data in Fig. 5 cover the energy range of approximately 0.6 eV to 1.0 eV where significant warping is observed in Fig. 6(c). In these warped contours the dominant scattering vector is $q_2$ connecting extremal parts of the Fermi surface (Fig. 6(c)) [35,37,38]. To determine the scattering vectors $q_2$ we fit the $dI/dV$ intensity oscillations to a power-law decay, $A\cos(q_2 x)/x^{3/2}$ [38], as shown in Fig. 6(d). The resulting $E$ vs. $q$ dispersion is plotted for two gate voltages in Fig. 6(e). The dispersions are offset from each due to the shift of the Dirac point with gate potential. We recover the $E$ vs. $k$ dispersion by connecting the scattering vectors to the momentum k. Recent calculations [38] have shown that $q_2$=1.5 $k$, where $k$ is measured from $\Gamma$. The resulting dispersion is shown in Fig. 7, where both the $V_G$ = 0 V and $V_G$ = -100 V data collapse onto a single curve when the energy is measured from the Dirac point. A linear fit to this data gives the local curvature of the $E$-$k$ dispersion of the surface states with a velocity of $(1.1\pm0.1) \times 10^6$ m/s. This velocity is almost twice the velocity determined at lower energies on thick films with a linear dispersion [38]. At this high energy above the Dirac point the dispersion is expected to be no longer linear due to warping and also due to the finite band gap [6], which may account for the increase in the local velocity determined from the QPI measurements.

In summary, we have successfully fabricated *in-situ* gate tunable epitaxial $Bi_2Se_3$ films. We have studied the effects of the gate's electric field as expressed in the shifts of spectral features in the tunneling spectra and in QPI scattering periods. While the current gating effectiveness is limited in $Bi_2Se_3$ due to bulk doping, the combination of *in-situ* MBE, STM, and transport studies in TI devices opens new avenues for future work in more insulating samples.

**Figure Captions**

Fig. 1. 3D computer-automated drawings of the high temperature sample holder with *in situ* back gating capability. (a) Top view. (b) Cross-sectional view. Part list: (1) alumina sample holder, (2) W source/drain electrode (tunnel bias), (3) W source/drain electrode, (4) $SrTiO_3$ substrate, (5) Pt electrodes on $SrTiO_3$, (6) Pt back gate electrode on $SrTiO_3$, (7) W spring clip to hold SrTiO3, (8)W back gate electrode.

Fig. 2. Characterization of the $SrTiO_3$ substrate for TI growth. (a) AFM image, 1 μm x 1 μm, of SrTiO3 after thermal processing as detailed in the text showing large atomically flat terraces separated by single atomic-height steps. (b) RHEED pattern of $SrTiO_3$ prior to TI growth. (c) RHEED pattern of 3 QL Bi2Se3 film grown on $SrTiO_3$. (d) STM topographic image, 81 nm x 81 nm, of 3 QL $Bi_2Se_3$ film grown by MBE on the gated sample holder in Fig. 1. The arrow indicates the position of *dI/dV* mapping of quasiparticle scattering in Fig. 5. Tunneling parameters: $V_B$=1.5 V, $I$=30 pA.

Fig. 3. Transport properties of 3 QL $Bi_2Se_3$ film. (a) $Bi_2Se_3$ film (3 QL) 2-terminal film resistance versus temperature. (b) $Bi_2Se_3$ film (3 QL) 2-terminal film resistance versus gate voltage.

Fig. 4. Electric field effect of $Bi_2Se_3$. (a) *dI/dV* spectra of 3 QL $Bi_2Se_3$ film versus gate voltage. The spectra are shifted vertically for clarity. The dashed lines are guides to the eye to indicate the approximate positions of the quantum well state peak (QW, black), the Dirac point (DP, purple), and the hump peak that defines the surface band gap (HP, green). Tunnel parameters: $I$=100 pA, $V_B$=0.3 V, $V_{mod}$=10 mV. (b) The shift of the leftmost QW peak versus gate voltage. The error bars are one standard deviation uncertainty in the peak position obtained by fitting the QW peak to a Lorentzian function. The solid line is a linear fit yielding the gating efficiency of (1.2±0.1) mV/V. The uncertainty in slope is one standard deviation uncertainty determined from the linear fit.

Fig. 5. $Bi_2Se_3$ quasiparticle interference (QPI) patterns from the step edge indicated in Fig. 2d. *dI/dV* maps of QPI patterns at $V_G$ = 0 V (a-d) and $V_G$ = -100 V (f-j) at the indicated sample biases. *dI/dV* intensity is given by the color scale from low (dark) to high (bright) in arbitrary units. The oscillations in the dI/dV maps are used to determine the energy momentum dispersion in Figs. 6 and 7. Tunneling parameters: $I$=50 pA, $V_B$ indicated in each panel, $V_{mod}$=15 mV.

Fig. 6. Quasiparticle scattering and LDOS oscillations from $Bi_2Se_3$ as a function of back gating. (a) Real space $Bi_2Se_3$ (111) surface lattice. Step edges occur along closed packed directions indicated along the $a_1$ unit cell vector. (b) Corresponding reciprocal space lattice and surface Brillouin zone (blue hexagon). Scattering at the step edges correspond to scattering in the Γ-M direction. (c) Fermi surface contours given by Eq. 1 for energies

relative to the Dirac point of 0.1 eV to 1.1 eV. Scattering oscillations are dominated by the vector $q_2$ connecting extremal parts of the Fermi contours along the Γ-M direction. (d) *dI/dV* intensity oscillations from scattering at the step edge in Fig. 2d obtained from the map in Fig. 5(d) at $V_B$=0.3 V and $V_G$=0 V. A smooth background was subtracted from the raw intensity to determine the oscillatory part (red data points), which was fit to a function $A\cos(qx)/x^{3/2}$ (blue line), where amplitude *A* and scattering vector *q* are fitting parameters. (e) Energy versus scattering vector obtained from *dI/dV* intensity oscillations as in (d) for gate voltages of 0 V and -100 V. The solid lines are linear fits to the data. The error bars are one standard deviation uncertainty in the scattering vector determined from fitting the intensity oscillations in (d).

Fig. 7. $Bi_2Se_3$ energy-momentum dispersion along the Γ-M direction determined from quasiparticle scattering oscillations from step edges. The *E* vs. *q* data in Fig. 6(e) at different gate voltages are repotted with energies relative to the Dirac point at $V_B$=-0.4 eV for $V_G$=0 V and $V_B$=-0.27 V for $V_G$=-100 V. The momentum *k* is determined from the scattering vector as q=1.5k following Ref. 35. The data from the different gate voltages collapse onto the same dispersion. The solid line is a linear fit yielding a velocity = (1.11±0.09) x $10^6$ m/s. The error in the velocity is one standard deviation uncertainty from the linear fit.

# References

* To whom correspondence should be addressed:joseph.stroscio@nist.gov

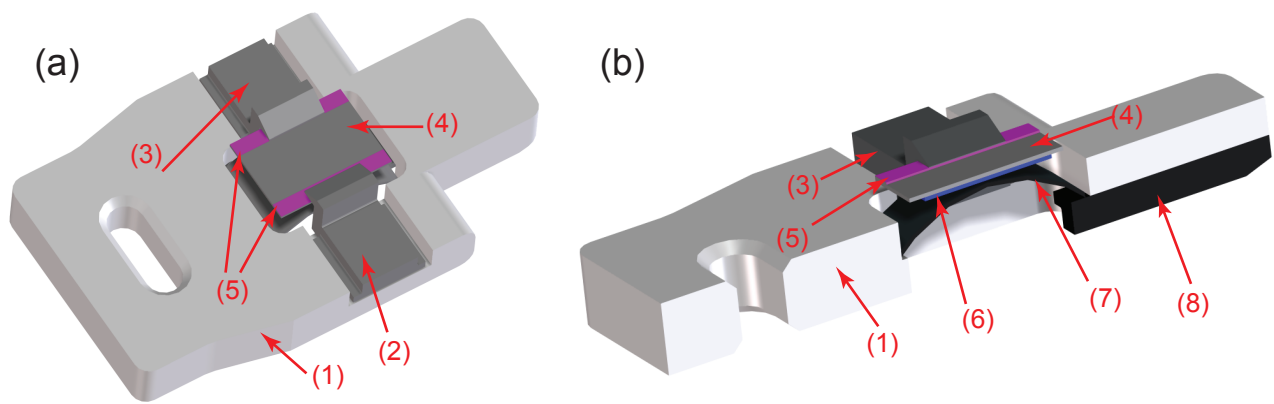

**Fig. 1**. 3D computer-automated drawings of the high temperature sample holder with *in situ* back gating capability. (a) Top view. (b) Cross-sectional view. Part list: (1) alumina sample holder, (2) W source/drain electrode (tunnel bias), (3) W source/drain electrode, (4) $SrTiO_3$ substrate, (5) Pt electrodes on $SrTiO_3$, (6) Pt back gate electrode on $SrTiO_3$, (7) W spring clip to hold $SrTiO_3$, (8) W back gate electrode.

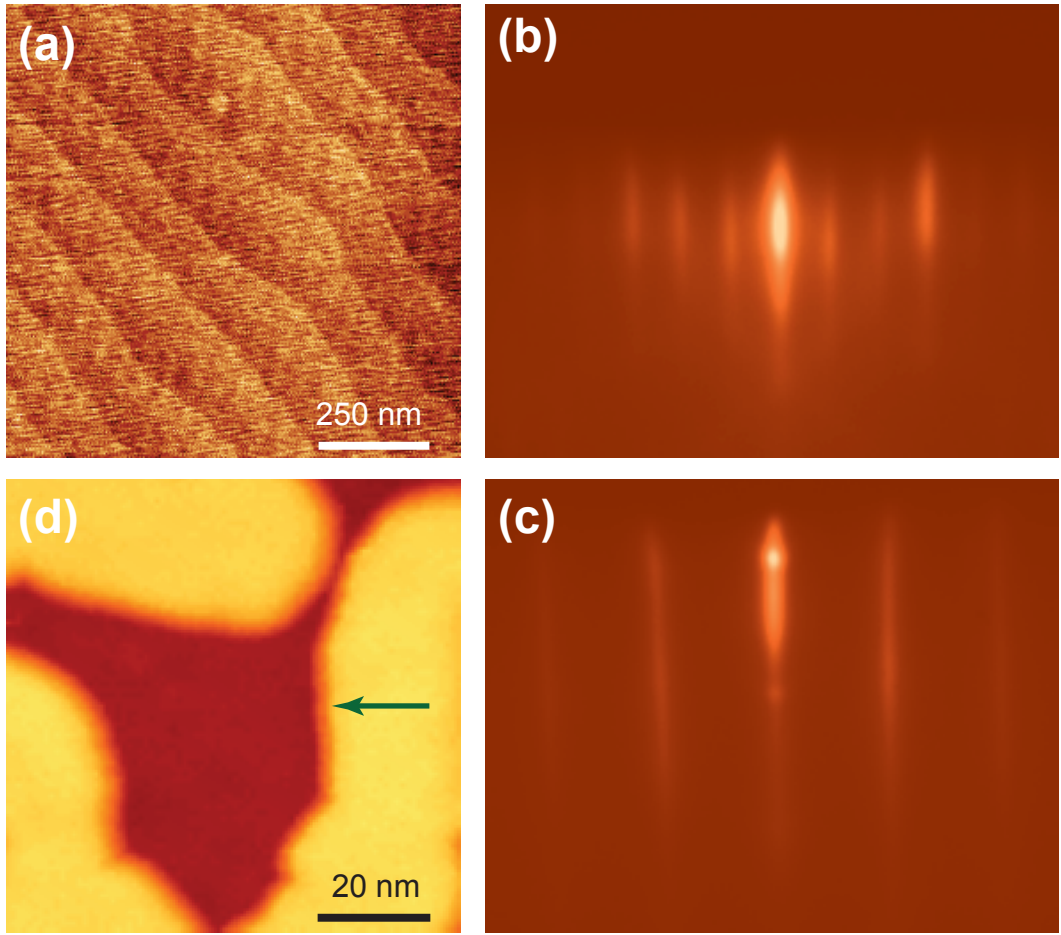

**Fig. 2**. Characterization of the SrTiO$_3$ substrate for TI growth. (a) AFM image, 1 μm x 1 μm, of SrTiO$_3$ after thermal processing as detailed in the text showing large atomically flat terraces separated by single atomic-height steps. (b) RHEED pattern of SrTiO$_3$ prior to TI growth. (c) RHEED pattern of 3 QL Bi$_2$Se$_3$ film grown on SrTiO$_3$. (d) STM topographic image, 81 nm x 81 nm, of 3 QL Bi$_2$Se$_3$ film grown by MBE on the gated sample holder in Fig. 1. The arrow indicates the position of *dI/dV* mapping of quasiparticle scattering in Fig. 5. Tunneling parameters: $V_B$=1.5 V, $I$=30 pA.

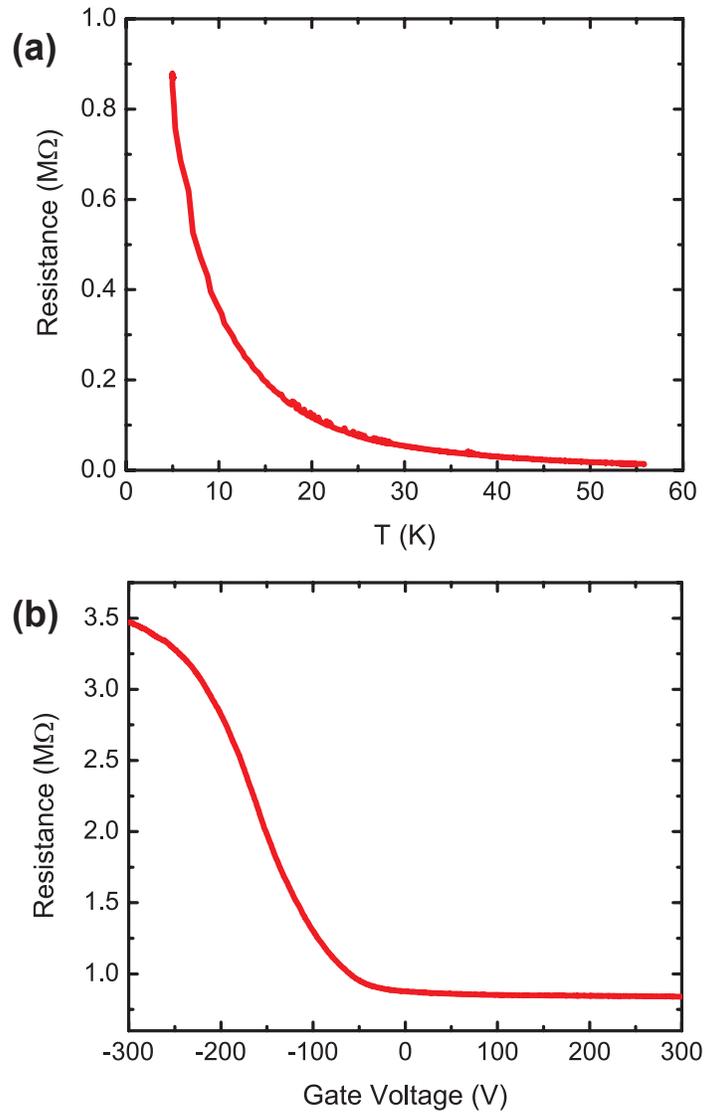

**Fig. 3.** Transport properties of 3 QL $Bi_2Se_3$ film. (a) $Bi_2Se_3$ film (3 QL) 2-terminal film resistance versus temperature. (b) $Bi_2Se_3$ film (3 QL) 2-terminal film resistance versus gate voltage.

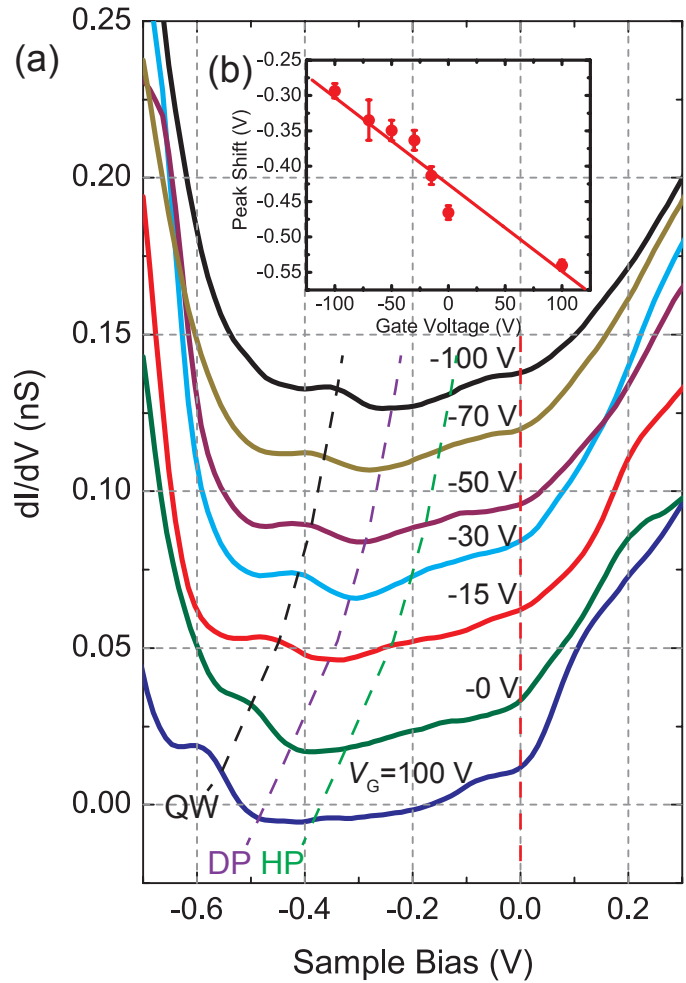

**Fig. 4**. Electric field effect of $Bi_2Se_3$. (a) $dI/dV$ spectra of 3 QL $Bi_2Se_3$ film versus gate voltage. The spectra are shifted vertically for clarity. The dashed lines are guides to the eye to indicate the approximate positions of the quantum well sate peak (QW, black), the Dirac point (DP, purple), and the hump peak that defines the surface band gap (HP, green). Tunnel parameters: $I$=100 pA, $V_B$=0.3 V, $V_{mod}$=10 mV. (b) The shift of the leftmost QW peak versus gate voltage. The error bars are one standard deviation uncertainty in the peak position obtained by fitting the QW peak to a Lorentzian function. The solid line is a linear fit yielding the gating efficiency of (1.2±0.1) mV/V. The uncertainty in slope is one standard deviation uncertainty determined from the linear fit.

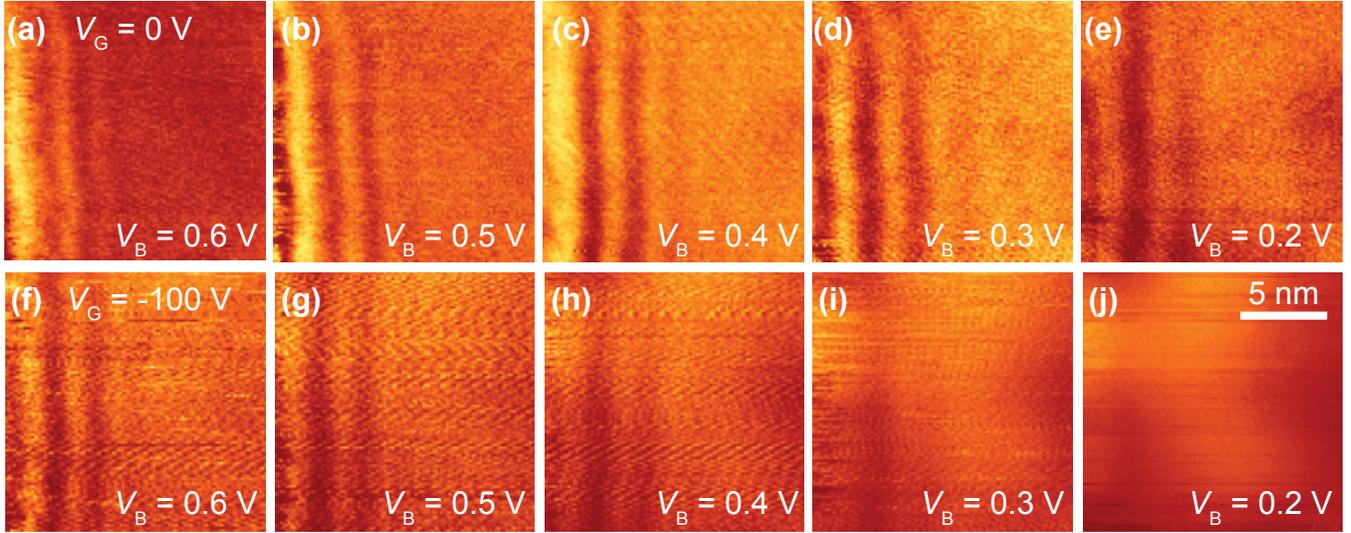

**Fig. 5**. Bi$_2$Se$_3$ quasiparticle interference (QPI) patterns from the step edge indicated in Fig. 2d. *dI/dV* maps of QPI patterns at $V_G$ = 0 V (a-d) and $V_G$ = -100 V (f-j) at the indicated sample biases. *dI/dV* intensity is given by the color scale from low (dark) to high (bright) in arbitrary units. The oscillations in the *dI/dV* maps are used to determine the energy momentum dispersion in Figs. 6 and 7. Tunneling parameters: *I*=50 pA, $V_B$ indicated in each panel, $V_{mod}$=15 mV.

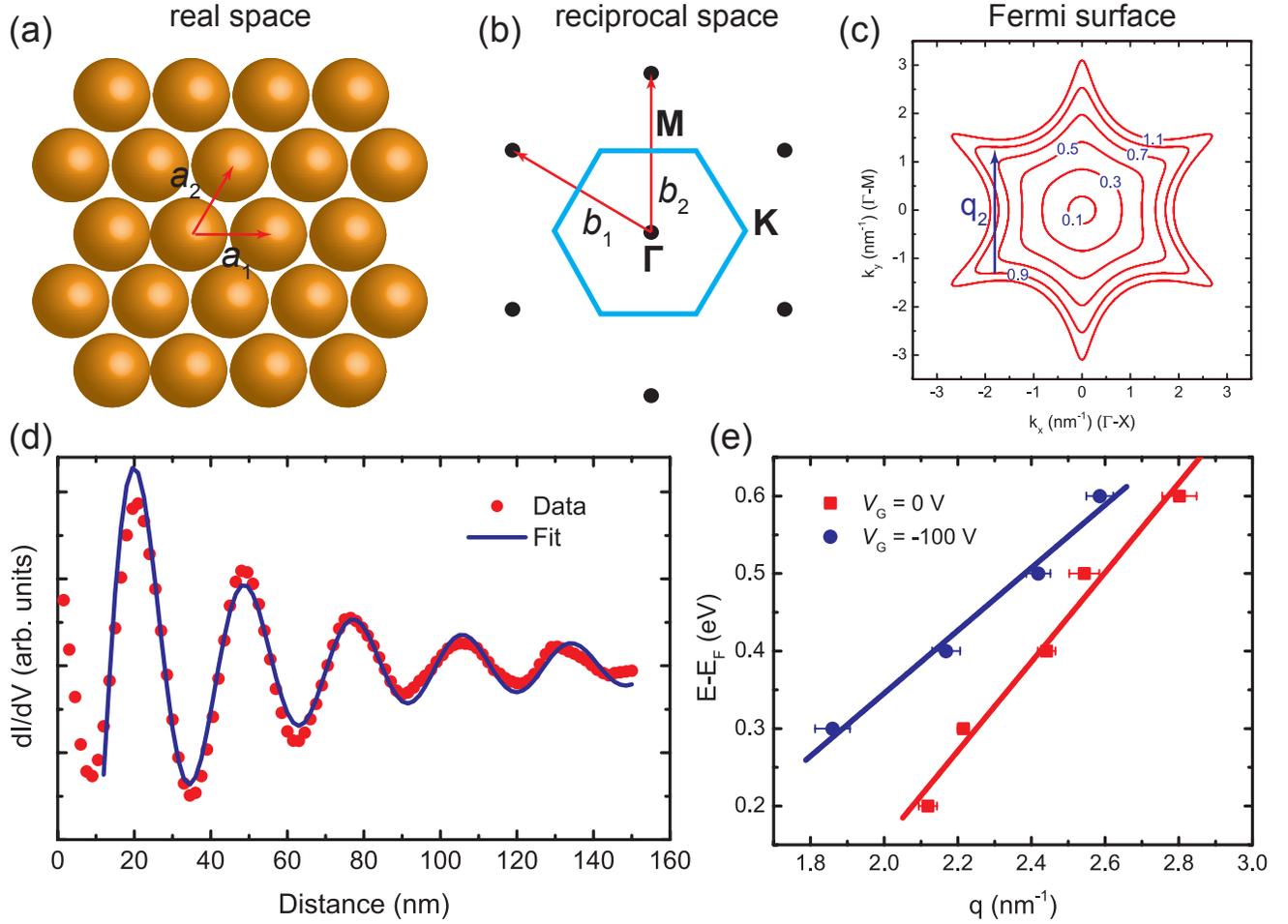

**Fig. 6**. Quasiparticle scattering and LDOS oscillations from $Bi_2Se_3$ as a function of back gating. (a) Real space $Bi_2Se_3$ (111) surface lattice. Step edges occur along closed packed directions indicated along the **a**$_1$ unit cell vector. (b) Corresponding reciprocal space lattice and surface Brillouin zone (blue hexagon). Scattering at the step edges correspond to scattering in the Γ-M direction. (c) Fermi surface contours given by Eq. 1 for energies relative to the Dirac point of 0.1 eV to 1.1 eV. Scattering oscillations are dominated by the vector $q_2$ connecting extremal parts of the Fermi contours along the Γ-M direction. (d) $dI/dV$ intensity oscillations from scattering at the step edge in Fig. 2d obtained from the map in Fig. 5d at $V_B$=0.3 V and $V_G$=0 V. A smooth background was subtracted from the raw intensity to determine the oscillatory part (red data points), which was fit to a function $A\ cos(qx)/x^{3/2}$ (blue line), where amplitude $A$ and scattering vector $q$ are fitting parameters. (e) Energy versus scattering vector obtained from $dI/dV$ intensity oscillations as in (d) for gate voltages of 0 V and -100 V. The solid lines are linear fit to the data. The error bars are one standard deviation uncertainty in the scattering vector determined from fitting the intensity oscillations in (d).

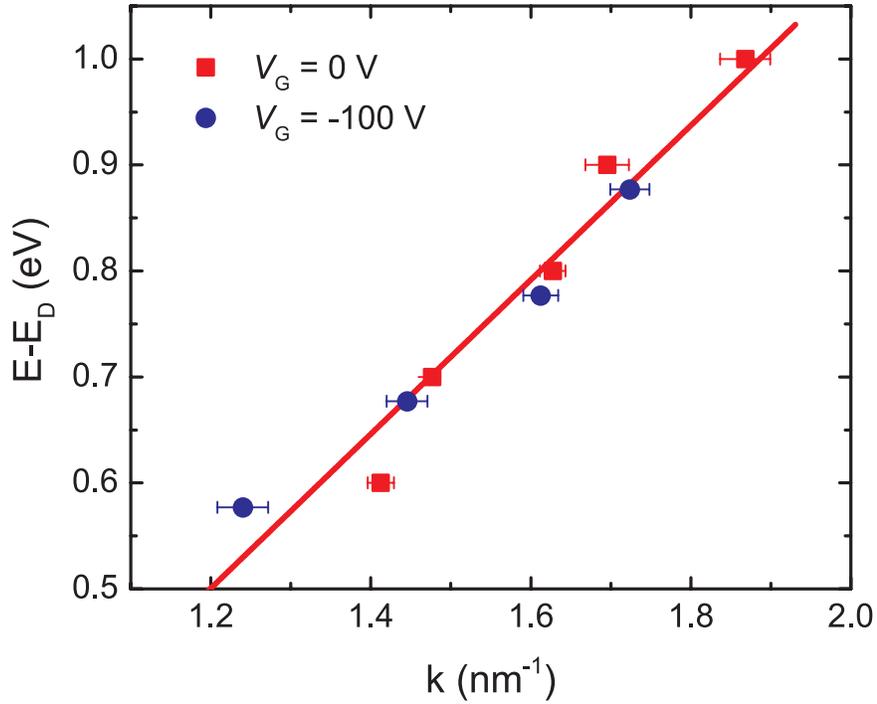

**Fig. 7**. $Bi_2Se_3$ energy-momentum dispersion along the Γ-M direction determined from quasiparticle scattering oscillations from step edges. The *E* vs. *q* data in Fig. 6e at different gate voltages are replotted with energies relative to the Dirac point at $V_B$=-0.4 eV for $V_G$=0 V and $V_B$=-0.27 V for $V_G$=-100 V. The momentum *k* is determined from the scattering vector as *q*=1.5*k* following Ref. 35. The data from the different gate voltages collapse onto the same dispersion. The solid line is a linear fit yielding a velocity = (1.11±0.09) x $10^6$ m/s. The error in the velocity is one standard deviation uncertainty from the linear fit.